# Statistical Properties of a Quantum Cellular Automaton

Norio Inui*

*Graduate School of Engineering,*

*University of Hyogo,*

*2167, Shosha, Himeji,*

*Hyogo, 671-2201, Japan*

Shuichi Inokuchi[†] and Yoshihiro Mizoguchi[‡]

*Faculty of Mathematics, Kyushu University,*

*6-10-1 Hakozaki, Higashi-ku,*

*Fukuoka, 812-8581, Japan*

Norio Konno[§]

*Department of Applied Mathematics,*

*Yokohama National University,*

*79-5 Tokiwadai, Yokohama,*

*240-8501, Japan*

(Dated: June 28, 2005)

# Abstract

We study a quantum cellular automaton (QCA) whose time-evolution is defined from global transition function of classical cellular automata (CA). In order to investigate natural transformations from CA to QCA, the present QCA includes CA with Wolfram's rule 150 and 105 as special cases. We firstly compute the time-evolution of the QCA and examine its statistical properties. As a basic statistical value, the probability of finding an active cell averaged over a spatial-temporal space is introduced, and the difference between CA and QCA is considered. In addition, it is shown that statistical properties in QCA are related to the classical trajectory in the configuration space.

*Electronic address: inui@eng.u-hyogo.ac.jp
†Electronic address: inokuchi@math.kyushu-u.ac.jp
‡Electronic address: ym@math.kyushu-u.ac.jp
§Electronic address: norio@mathlab.sci.ynu.ac.jp

# I. INTRODUCTION

Classical cellular automata (CA) have played an important role in classical complex systems. Thus the quantization of CA is a fascinating challenge[1]. In particular, the study of the quantum cellular automata (QCA) is useful for analysis of quantum computers. Because John Watrous proved that a certain QCA has the same potentiality with the quantum Turing machine [2]. Although implementation of QCA is not satisfactorily realized, quantum-dot CA is a promising candidate[3, 4].

The rule of CA is deterministic and very simple, but it exhibits often very complex behavior. Therefore CA is a good model to study a bridge between statistical mechanics and thermodynamics[5]. The rule of QCA introduced in this paper is also simple, but analysis of its behavior is not easy. We consider the statistical properties of QCA by taking average in a spatial-temporal space and show universal properties which are independent of the system size and initial states.

The time evolution of QCA has not been widely discussed as compared with that of CA. There are two major reasons. First a configuration in CA is mapped to a single configuration at each step. Thus the simulation is easy even if the system size is rather large. In contrast the simulation of QCA by using current computers is extremely difficult for large system size. Because a configuration transits superimposition of many different configurations. Second the act of observation changes the state of QCA. The observation breaks a superposition of configuration, so that the state of QCA will be a single configuration after observation. If this observation is carried out at each step, we must take account of the Zenon effect.

In this paper, we consider QCA based on the following CA with two nearest-neighbor sites. Each cell is arranged on a circle and it has one of two possible states 0(inactive) and 1(active). We denote the state of a cell indexed by an integer $k$ as $q_k \in \{0,1\}$. If the number of cells is $K$, then a configuration is defined by a set $\{q_0, q_1, \ldots, q_{K-1}\}$ and it is simply denoted by $\xi_i$ with an integer $i \equiv \sum_{k=0}^{K-1} q_k 2^k$. The state of a cell at position $k$ is determined from the states $q_k, q_{k-1}$ and $q_{k+1}$. In particular, the local transition function of CA with Wolfram's rule 150 is given by $(q_{k-1} + q_k + q_{k+1}) \mod 2$.

To connect CA with QCA, it is useful to express the time evolution of CA using a matrix. We introduce a state vector $\phi(t)$ whose $i$-th element is 1 if the configuration of CA is $\xi_i$ at time $t$ and 0 otherwise. We furthermore define a transition matrix $M$ whose element in row $i$ of column $j$ is 1 if the transition from a configuration $\xi_j$ to a configuration $\xi_i$ is permitted and 0 otherwise. Using this matrix the state vector at $t+1$ is obtained by $M\phi(t)$. Since the dynamics of CA is deterministic, we surely observe only one configuration at fixed time. A quantum state of QCA, by contrast, is superposition of configurations, and we can calculate only the probability amplitude associated with each configuration in superposition. Similarly with CA, the probability amplitude is also calculated by operating the time-evolution matrix to the quantum state vector. The significant difference in time-evolution matrix between CA and QCA is that the matrix corresponding to QCA must be unitary. Therefore it is not obvious whether there is a simple way to define QCA from CA or not. Recently a simply way to extend CA to QCA is introduced[6], however, the details has been untouched. Thus the aim of this paper is simulating their QCA whose time evolution is defined using Wolfram's rule 150 and considering the statistical properties.

The remainder of this paper will be organized as follows. In Sec. II, a simple method of extension from CA to QCA is introduced. In Sec. III, the mean number of active cells averaged over time and space is considered. In Sec. IV, the time dependence of the probability of finding a given configuration is calculated for different system size, and it is shown that the configuration appearing in the dynamics of CA is observed with high probability. Finally, in Sec. V, the difference between CA and QCA is mentioned.

## II. EVOLUTION OPERATOR OF QCA

As we mentioned in the introduction, the probability amplitude of a configuration $\xi_i$ is essential to analysis of QCA. There are $N \equiv 2^K$ configurations in CA with system size $K$, therefore, the time-evolution of a quantum state in $N$ dimensional Hilbert space is considered. Let us denote the state vector of QCA at time $t$ as $\phi(t)$ whose $i$-th element is amplitude associated with the configuration $\xi_i$. The probability of finding a configuration $\xi_i$ is the absolute square of the amplitude. The state vector of QCA is updated by operating evolution matrix $M$. The element of $M$ in row $i$ of column $j$ denotes the amplitude of transition from a configuration $\xi_j$ to a configuration $\xi_i$. The rule of QCA is completely determined by this matrix and an initial state $\phi_0 \equiv \phi(0)$.

We proposed a new formulation of a finite cyclic QCA by a natural transformation from CA [6]. First of all a local transition function in CA must be defined. Let a state of a cell indexed by $k$ in CA be $q_k$. The value $q_k$ in CA with a neighborhood size 1 is updated by a local map $f(q_{k-1}, q_k, q_{k+1})$. In this paper we use the following function:

$$f(q_{k-1}, q_k, q_{k+1}) = q_{k-1} + q_k + q_{k+1} \pmod{2}. \tag{1}$$

This is a local transition function of CA governed by the Wolfram's rule 150. Since the finite cells are located cyclically we identity $q(-1)$ with $q(K-1)$ and $q(K)$ with $q(0)$ in the above function.

We now turn to the definition of the time-evolution matrix in QCA. Let us denote the $k$-th element of a given configuration $\xi_i$ by $\xi_i(k)$. Then the element of matrix $M_{ij}$ is defined by a product of a local function which is determined by a tripartite $\{\xi_j(k-1), \xi_j(k), \xi_j(k+1)\}$ and $\xi_i(k)$. Taking account into periodic boundary condition the matrix element $M_{ij}$ is defined by

$$M_{ij} = \prod_{k=0}^{K-1} g(\xi_j(k-1), \xi_j(k), \xi_j(k+1); \xi_i(k)), \tag{2}$$

where

$$g(q_{k-1}, q_k, q_{k+1}; p_k) = \begin{cases} \cos\theta, & \text{if } f(q_{k-1}, q_k, q_{k+1}) = 0 \text{ and } p_k = 0, \\ -\sin\theta, & \text{if } f(q_{k-1}, q_k, q_{k+1}) = 0 \text{ and } p_k = 1, \\ \sin\theta, & \text{if } f(q_{k-1}, q_k, q_{k+1}) = 1 \text{ and } p_k = 0, \\ \cos\theta, & \text{if } f(q_{k-1}, q_k, q_{k+1}) = 1 \text{ and } p_k = 1. \end{cases} \tag{3}$$

The principle of quantum mechanics requires that the matrix $M$ is a unitary matrix. We can show that the above matrix is a unitary matrix if and only if the number of cells is not multiples of the number 3 (see details in Appendix A). Thus we assume that the size $K$ is not multiples of the number 3 in following sections.

Let $\delta_i$ be a unit vector whose $i$-th element is 1. If the parameter $\theta$ is zero, the matrix $M$ is the transition matrix of CA with Wolfram's rule 150, and the position of 1 in the state vector $M^t \delta_i$ shows the configuration of CA after $t$ step starting from the configuration $\delta_i$. Thus the above QCA includes CA with Wolfram's rule 150 as a special case.

Let us calculate the matrix element $M_{53}$ for $K = 4$ as an example. This element denotes the amplitude of transition from a configuration $\xi_3 = \{0, 0, 1, 1\}$ to $\xi_5 = \{0, 1, 0, 1\}$. The configuration $\xi_3$ transits to $\xi_{12} = \{1, 1, 0, 0\}$ by mapping of CA with Wolfram's rule 150. From the definition the value of $M_{53}$ is given by $g(1, 0, 0; 1)g(0, 0, 1; 1)g(0, 1, 1; 0)g(1, 1, 0; 0) =$ $- \cos^2 \theta \sin^2 \theta$. Similarly we have $M_{02} = \cos \theta \sin^3 \theta$ and $M_{00} = \cos^4 \theta$ for other examples. To calculate more simply we define a humming distance between $\xi_i$ and $\xi_j$ by

$$d(\xi_i, \xi_j) = \sum_{k=0}^{K-1} (\xi_i(k) - f(\xi_j(k-1), \xi_j(k), \xi_j(k+1)). \tag{4}$$

The humming distance takes a value between 0 and $K-1$. If the configuration $\xi_i$ is obtained by mapping from $\xi_j$ in CA, the humming distance is 0. Using this humming distance we express the absolute value of $M_{ij}$ by

$$|M_{ij}| = \cos^{K-d(\xi_i,\xi_j)} \theta \sin^{d(\xi_i,\xi_j)} \theta. \tag{5}$$

If the value of $\theta$ is between 0 and $\pi/2$, the absolute amplitude of $M_{ij}$ is larger as the hamming distance is small. Suppose that the initial state is $\delta_i$. Then the configuration obtained by a single CA mapping is observed with high probability for small $\theta$. For this reason, we may say that the magnitude of $\theta$ represents the strength of quantum effect.

As the first step to characterize the evolution of QCA we consider the probability of finding an active state in a cell indexed by $k$ at time $t$. This probability depends on not only time but also the previous observation. Assuming that the observation is firstly performed at time $t$, the probability starting from initial state $\phi_0$ is given by

$$P(k, t, \phi_0) = \sum_{i=0}^{N-1} \xi_i(k) |\phi_i(t, \phi_0)|^2, \tag{6}$$

where $\phi_i(t, \phi_0)$ is the probability amplitude corresponding to a configuration $\xi_i$ starting from an initial state $\phi_0$.

The graphical representation of active cells in space-time assists greatly our understanding. We show the probability $P(k, t, \phi_0)$ of QCA with five cells starting from an initial state $\delta_4$ in Fig.1. The amplitude of the probability is shown by light and shade. The probability is proportion to the shade level. If the parameter $\theta$ is zero, the probability of each cell takes either 0 or 1. Thus the black square denotes active cell in CA with Wolfram's rule 150 and we find that the configuration changes with a period 3. The change of configuration corresponding to $\theta = \pi/16$ is similar to that of CA for small number of steps. However, active cells are observed at positions where the active cell is never observed in the case of $\theta = 0$. When the parameter increases, the deviation of the light level decreases and a pattern with a period 4 appeasers at $\theta = \pi/4$. We stress here that QCA with the parameter $\theta = \pi/4$ is not classical cellar automaton. All configurations are observed with the same probability at $t = 1, 3, 5, \ldots$. One find a classic pattern at $\theta = \pi/2$ again. This is the same with a pattern seen in CA with Wolfram's rule $105 = (2^8 - 1) - 150$. As a result, we find that the parameter $\theta$ connects between CA with Wolfram's rule 150 and CA with Wolfram's rule 105. The structure of the sketches near $\theta = 0$ correlates with the pattern of CA with Wolfram's rule 150 and the structure of the sketches near $\theta = \pi/2$ correlates with the pattern of CA with Wolfram's rule 105.

## III. DEFINITION OF MEAN DENSITY OF ACTIVE CELLS

We begin to consider the density of active cells at fixed time as a basic statistical value. In contrast with CA, it is not obvious that we can measure the state of every cell at the same time. Thus we assume that we can choose a cell randomly at fixed time and observe the state of it at least. We repeat this measurement and count the number of the occurrence of observing an active cell. The ratio of this number to the total number of measurements is regarded as mean density of active cell. The ratio as the number of measurements tends to infinity converges to

$$\rho(t, \theta, \phi_0, K) = \frac{1}{K} \sum_{k=0}^{K-1} P(k, t, \phi_0). \tag{7}$$

We show the time dependence of the mean density of active cells $\rho(t, \theta, \phi_0, K)$ for $\theta = 0.35764$ and $K = 5$ as an example in Fig. 2. The initial states of Fig. 2(a) and Fig. 2(b) is $\delta_0$ and $\delta_{11}$, respectively. The values of $\rho(t, \theta, \phi_0, K)$ always fluctuates except special configurations discussed in the last of this paper. For this reason, we consider the time-averaged mean density. Let us define generally the time-averaged value of a time-dependent function $f(t)$ by

$$\langle f(t, \phi_0) \rangle \equiv \lim_{T \to \infty} \frac{1}{T} \sum_{t=0}^{T-1} f(t, \phi_0). \tag{8}$$

From this definition the time-averaged mean density is given by

$$\rho(\theta, \phi_0, K) = \langle \rho(t, \theta, \phi_0, K) \rangle. \tag{9}$$

The averaged values of data shown in Fig. 2(a) and Fig. 2(b) are 0.4945 and 0.4997. These values are indicated by horizontal lines. Although it is impossible to obtain the exact values of $\rho(\theta, \phi_0, K)$ by simulations, both values are very close to 1/2. Indeed, we will show that the value of $\rho(\theta, \phi_0, K)$ is exactly 1/2 independently on the initial sates in the following sections. From this universal property it follows that two time evolution of QCA are not distinguished by the mean density of active cells, but the amplitude of fluctuations are clearly different. To characterize these fluctuations we introduce the following time-averaged standard deviation:

$$\sigma(\theta, \phi_0, K) = \left\langle (\rho(t, \theta, \phi_0, K) - \rho(\theta, \phi_0, K))^2 \right\rangle^{1/2}. \tag{10}$$

The deviations corresponding to data in Fig. 2(a) and Fig. 2(b) are 0.191 and 0.088.

## IV. EXACT SOLUTIONS FOR $K = 4$

If all eigenvalues and all eigenvectors of the time evolution matrix $M$ are obtained, all quantum states of QCA are exactly calculated for any initial states. This may be hard task for large system size, but eigenvalues and eigenvectors of $M$ with $K = 4$ are easily computed, and the time-averaged probability of finding a configuration $\xi_i$, i.e. $\langle |\phi_i(t, \phi(0))|^2 \rangle$ is exactly calculated. The problem whether observing every state at once is possible or not is ignored for the purpose of this paper. We show only the results starting initial states $\delta_0 = \{0, 0, 0, 0\}$ and $\delta_3 = \{0, 0, 1, 1\}$ for $0 < \theta < \pi/2$:

$$\left\langle |\phi_i(t, \delta_0)|^2 \right\rangle = \begin{cases} \frac{83}{256} + \frac{3}{64} \cos 4\theta + \frac{1}{256} \cos 8\theta, & i = 0, 15, \\ \frac{1}{32} + \frac{1}{128} \sin^2 4\theta, & i = 1, 2, 4, 7, 8, 11, 13, 14, \\ \frac{1}{32} \sin^4 2\theta, & i = 3, 5, 6, 9, 10, 12, \end{cases} \tag{11}$$

and

$$\left\langle |\phi_i(t, \delta_3)|^2 \right\rangle = \begin{cases} \frac{1}{32} + \frac{1}{128} \sin^2 4\theta, & i = 0, 3, 5, 6, 9, 10, 12, 15, \\ \frac{1}{32} \sin^4 2\theta, & i = 1, 4, 11, 14, \\ \frac{51}{256} - \frac{1}{64} \cos 4\theta + \frac{1}{256} \cos 8\theta, & i = 2, 13, \\ \frac{35}{256} + \frac{3}{64} \cos 4\theta + \frac{1}{256} \cos 8\theta. & i = 7, 8. \end{cases} \tag{12}$$

A further important point is that time-average probability of $\xi_i$ for $i = 0, 1, \ldots, 7$ is the same with the configuration $\xi_{15-i}$ that is obtained by reversing each bit in a configuration $\xi_i$. From this relation, we can conclude immediately that both of $\rho(\theta, \delta_0, 4)$ and $\rho(\theta, \delta_3, 4)$ are exactly a half. Since the relation between a configuration $\xi_i$ and $\xi_{(N-1)-i}$ is often referred in this paper, we denote $\xi_{(N-1)-i}$ by $\hat{\xi}_i$.

To obtain the time-averaged standard deviations some arithmetic are required, but they are straightforward given by

$$\sigma(\theta, \delta_0, 4) = \sqrt{\frac{13 + 3\cos 4\theta}{128}}, \tag{13}$$

$$\sigma(\theta, \delta_3, 4) = \sqrt{\frac{13 + 3\cos 4\theta}{512}}, \tag{14}$$

where $0 < \theta < \pi/2$. In the both cases the time-averaged standard deviation takes a minimum value at $\theta = \pi/4$.

Let us compare these results with values for classical case. The number of active cells of CA with Wolfram's rule 150 starting from $\delta_0$ is always 0. Thus both of the mean number of active cells and the standard deviation are also 0. On the other hand $\lim_{\theta \to +0} \rho(\theta, \delta_0, K) = 1/2$ and $\lim_{\theta \to +0} \sigma(\theta, \delta_0, 4) = 1/(2\sqrt{2})$. This means that there is no smooth crossover for both of the mean number of active cells and standard deviation between CA and QCA.

## V.  EXACT TIME-AVERAGED MEAN DENSITY OF ACTIVE CELLS

We show that the time-averaged mean density of active cells of QCA for $0 < \theta < \pi/2$ is exactly $1/2$. Put simply, this implies that a half of many cells chosen in space-time at random are observed in active state independently of initial states. As shown in the Sect. IV, we can conclude that the mean density of active cells of QCA is $1/2$ if the following relation is satisfied for all configurations

$$\langle |\phi_i(t, \phi_0)|^2 \rangle = \langle |\phi_{\hat{i}}(t, \phi_0)|^2 \rangle, \tag{15}$$

where $\hat{i} = (N - 1) - i$.

Since the time-evolution matrix $M$ is unitary, we have a diagonal matrix by a similarity transformation $V^\dagger M V$. Let $\lambda_n$ and $v_{ij}$ be an eigenvalue of $M$ and the element of $V$ in row $i$ of column $j$, respectively. The amplitude of the quantum state corresponding to a configuration $\xi_i$ is given by combination of $\lambda_n^t$:

$$\phi_i(t, \phi_0) = \sum_{n=0}^{N-1} \sum_{m=0}^{N-1} v_{i,n} v_{m,n}^* \phi_0(m) \lambda_n^t, \tag{16}$$

where $\phi_0(m)$ denotes the initial probability amplitude of the configuration $m$.

If there are degenerate eigenvalues of the matrix $M$, let us express $\phi_i(t, \phi_0)$ by the summation over different eigenvalues $\lambda_n (n = 0, 1, \ldots, l \leq N)$ and denote the coefficient of $\lambda_n$ in $\phi_i(t, \phi_0)$ by $a_{in}(\phi_0)$. The function $\phi_i(t, \phi_0)$ is expressed by

$$\phi_i(t, \phi_0) = \sum_{n=0}^{l} a_{in}(\phi_0) \lambda_n^t. \tag{17}$$

Thus the time-averaged probability $\langle |\phi_i(t, \phi_0)|^2 \rangle$ is given by

$$\langle |\phi_i(t, \phi_0)|^2 \rangle = \sum_{n=0}^{l} \sum_{m=0}^{l} a_{in}(\phi_0)(a_{im}(\phi_0))^* \langle \lambda_n^t (\lambda_m^t)^* \rangle. \tag{18}$$

Since the matrix $M$ is unitary, each eigenvalue $\lambda_n$ is written in the form of $e^{i\varphi_n}$. Using this expression we find easily

$$\langle \lambda_n^t (\lambda_m^t)^* \rangle = \begin{cases} 1, & \lambda_n = \lambda_m, \\ 0, & \lambda_n \neq \lambda_m. \end{cases} \tag{19}$$

Therefore function $\langle \phi_i(t, \phi_0) \rangle$ is given by

$$\langle |\phi_i(t, \phi_0)|^2 \rangle = \sum_{n=0}^{l} |a_{in}(\phi_0)|^2, \tag{20}$$

We here use a particular relation between elements in the matrix $M$:

$$M_{\hat{n}\hat{m}} = (-1)^{\sharp_n + \sharp_m} M_{nm}, \tag{21}$$

where $\sharp_i$ denotes a number of 1 of a decimal number $i$ in binary form, e.g. $\sharp_{11} = 3$. From this relation, we can show that there are eigenvectors whose elements satisfying

$$|v_{ij}| = |v_{\hat{i}j}|, \tag{22}$$

and they are $N$-dimensional orthonormal base. This means that the absolute value of $i$-th element of an eigenvector of $M$ is the same with the absolute value of $\hat{i}$-th element of the eigenvector (a sketch of proof of (22) is given in Appendix). From (22) it follows that $|a_{in}| = |a_{\hat{i}n}|$ in (20). As a consequence (15) is satisfied and we find that the mean density of active cells of QCA is $1/2$.

Let us regard an active cell in QCA as a particle. The number of particles changes depending on the configuration. First of all we probably ask the number of particle at fixed time. We can calculate its probability in principle, however, it is impossible to calculate by the current computer for large system size. In such a situation we often interest the statistical properties and we ask the mean number of particle per a site in thermodynamics limit. In QCA there is no stationary solution, therefore, we can not consider equilibrium states unlike the classical dynamics. For this reason we introduced the time-averaged mean density. The last conclusion that claming that time-averaged mean density is 1/2 holds for any system size except multiples of the number 3, and it means that there is a symmetry for replace "1" with "0" by taking average over time.

## VI. PROBABILITY DISTRIBUTION OF CONFIGURATIONS

The time-averaged one-point function was considered in the previous section. We here assume that a configuration of QCA is determined by a single observation and consider the time-averaged probability of finding the configuration. Since the probability of finding a configuration $\xi_i$ at time $t$ starting an initial state $\phi_0$ is $|\phi_i(t, \phi_0)|^2$, the time-averaged value is given by $\langle |\phi_i(t, \phi_0)|^2 \rangle$. Although the distribution strongly depends on both $\theta$ and initial states we see common properties in simulations. Figure. 3 shows simulation results for $\langle |\phi_i(t, \phi_0)|^2 \rangle$ at $\theta = 0.7$ as an example. The system size $K$ is $5, 7, 8, 10$ from top to bottom. The initial state is $\xi_{11} = \{1, 0, 1, 1\}$ independently of the system size. The number of steps is $N$ (3200 steps only $K = 5$). First of all we can confirm a mirror symmetry with respect to a vertical center line, which results from the symmetry $\langle |\phi_i(t, \phi_0)|^2 \rangle = \langle |\phi_{\bar{i}}(t, \phi_0)|^2 \rangle$. Second if the system size is large and $\theta \neq 0, \pi/4, \pi/2$, then the time-averaged probability for fixed $K$ is almost the same except several configurations. The exceptional large values are observed at the configurations that appear in CA with the same initial state. For example, if the system size $K$ is 5 and the initial state is $\delta_{11}$, the number of configuration $i$ changes periodically : $11 \to 8 \to 28 \to 11 \to \cdots$. The sequence corresponding to each system size is described in the figure. From the result in the previous section, the configurations obtained by reversing every bits in $\{11, 8, 28\}$, i.e., $\{20, 23, 3\}$ are also observed with high probabilities. Similarly the configuration in CA with $K = 7$ changes: $11 \to 88 \to 69 \to 44 \to 98 \to 22 \to 49 \to \cdots$. Therefore we find configurations $\{11, 22, 29, 39, 44, 49, 58, 69, 78, 83, 88, 98, 105, 116\}$ with high probabilities. However, we can not say that these configurations are observed with larger probability than other configurations. To take an example, the second largest value corresponding to $K = 7$ in Fig.3 is observed at the configuration $\xi_{35} = \{0, 1, 0, 0, 0, 1, 1\}$, which is similar with $\xi_{11} = \{0,0,0,1,0,1,1\}$.

Let us compare QCA with a stochastic CA in which time-evolution matrix $U$ is defined by replacing $M_{ij}$ by $|M_{ij}|^2$. The element of $U$ in row $i$ of column $j$ denotes a transition probability configuration from $\xi_j$ to $\xi_i$. If we regard $\phi_i(t)$ as a probability of finding a configuration $\xi_i$ of the stochastic CA, then the series obtained by multiplying the matrix $U$ to $\delta_i$ is a Markov process starting from an initial state $\delta_i$. The stationary distribution of probability of the stochastic CA is homogenous and it is different from Fig.3. Figure. 3 seems to have both characters of the stochastic CA and the deterministic CA.

We turn now to the time dependence of the probability of finding the configuration $\xi_i$. As we mentioned in Sec. II if the value of $\theta$ is small, the time-evolution of QCA is almost the same with QCA and CA for small $t$. However, long-time behaviors are quite different. For example, if the initial state is $\xi_0 = \{0,0,\ldots,0\}$, then the configuration $\xi_0$ is only observed in CA. Meanwhile the result in the previous section requires that the configuration $\hat{\xi}_0 = \{1,1,\ldots,1\}$ should be observed after long time. Let us estimate the time when the probability of finding the configuration $\hat{\xi}_0$ is larger than the probability of finding the configuration $\xi_0$ for QCA with $K = 4$ starting from $\xi_0$. From the exact solutions in Sect. IV the time is approximately given by $\pi/(2\theta)$. Thus the transition time becomes much longer as $\theta$ decreases.

To observe this reversal we show the change of probability of finding configurations $\xi_i$ in QCA with $K = 7$ and $\theta = 0.01$ in Fig. 4(a). The darkness of points is proportion to the probability $|\phi_i(t, \phi_0)|^2$. In Fig. 4(b) the order of configuration is arranged so that the temporal correlation is strong between two adjacent configurations. Furthermore the configurations in the right half is bit-reversal of the configuration in the left half. We can see slow a transition between configurations in a bit-reversal relation.

## VII. CONCLUSIONS

Simulations of QCA by current architecture need huge calculation time in comparison with CA. The computational complexity of CA with size $K$ increases in $\mathcal{O}(K)$, but that of QCA with size $K$ increases in $\mathcal{O}(2^K)$. Therefore analytical results are very valuable to study QCA with large system size. We introduced time-averaged density of active cells and showed that its value is exactly 1/2 independently of the system size and initial states. This result is derived from the symmetry $\langle |\phi_i(t,\phi_0)|^2 \rangle = \langle |\phi_{\hat{i}}(t,\phi_0)|^2 \rangle$. This universal property is very simple, but we can obtain several non-trivial results from this property. For example, this relation restricts stationary probability distributions. There is no stationary wavefunction in QCA. However there are many different stationary probability distributions. These distributions are more precisely defined by the distribution satisfying $|\phi_i(t,\phi_0)|^2 = |\phi_i(t+1,\phi_0)|^2$ for any $i$ and $t$. If we can choose initial state satisfying $|\phi_i(0,\phi_0)|^2 = |\phi_i(1,\phi_0)|^2$ for any $i$, then we always find a given configuration with the same probability. From the equation $\langle |\phi_i(t,\phi_0)|^2 \rangle = \langle |\phi_{\hat{i}}(t,\phi_0)|^2 \rangle$, the absolute value of such wave-function corresponding to a configuration $\xi_i$ must be the same with that corresponding to a configuration $\hat{\xi}_i$

We know little about the relation between CA and QCA. Our definition of QCA is very simple and it is the same with CA with Wolfram's 150 by setting $\theta = 0$. Therefore if the initial state is the same, the time-evolution near $\theta = 0$ is similar each other for small $t$. The difference, however, grows as time increases. This dose not mean the memory of CA is perfectly lost. Because configurations which appear in CA are observed with larger probability than other configurations. If we assume that the initial state of QCA is $\delta_i$, the time-averaged probability of finding probability $\xi_i$ is larger than other configuration except $\hat{\xi}_i$ in our simulations. We can not prove it, but if it is true, the memory of the initial state remains after time-averaging and we can infer the initial state from time-averaged probabilities.

We made clear several statistical properties of QCA, but we left some problems. First, the realization of QCA introduced here is the most important problem. As mentioned in many literatures, the restriction by "No-Go lemma" is a major obstacle to realize QCA. To overcome this difficulty, Karafyllidis proposed QCA with two qubits per cell [7]. One of qubits is the controlled qubit and the other is the state qubit. In his QCA a quantum controlled-NOT gate and the Hadamard gate are used. It is not obvious that our QCA is realized in his scheme, but the realization of QCA using quantum gates may be the most straightforward. Brennen and Williams studied another possibility [8]. In their QCA, the Hamiltonian is explicitly given, and it is a kind of a one-dimensional Ising spin. Thus the implementation of QCA can be considered more realistically. Since there is huge knowledge about Ising spin systems, the mapping of QCA to spin systems is very valuable. They analyzed information transport and entanglement of QCA. Their results gave us to extend QCA to open quantum systems. If we can regard the dynamics of QCA as a quantum process in non-equilibrium open systems, then physical quantities such as entropy may be used to describe the statistical properties of QCA. Second, we restricted our interests to QCA with Wolfram's rule 150 only. Thus we do not know yet statistical properties of QCA with other rules. Although extensive calculations are required to examine other QCA in comprehensive, our preliminary simulation suggests that the time-averaged probability of QCA dose not always converges 1/2. We conjecture that a certain symmetry of the Wolfram's rule determines whether $\rho = 1/2$ or not.

## APPENDIX A

We show that the matrix $M$ whose elements are defined by (2) is a unitary matrix if and only if $K \neq 0 \pmod{3}$. From the previous result [6], if the mapping of CA is a bijection then we can conclude that the time-evolution matrix of the QCA corresponding to the CA is unitary. Thus we prove that the mapping of CA with Wolfram's rule 150 is a bijection if and only if the system size $K$ is not multiples of the number 3.

Let a configuration in CA be $v \equiv \{q_0, q_1, \ldots, q_{K-1}\}^T$ where $T$ denotes a transpose operator. The mapping in CA with Wolfram's rule 150 is expressed by a mapping $f(v) = Av$ (mod 2) with the following matrix

$$A = \begin{bmatrix} 1 & 1 & 0 & 0 & 0 & \cdots & 0 & 0 & 1 \\ 0 & 1 & 1 & 1 & 0 & \cdots & 0 & 0 & 0 \\ \vdots & \vdots & \vdots & \vdots & \vdots & & \vdots & \vdots & \vdots \\ 0 & 0 & 0 & 0 & 0 & \cdots & 1 & 1 & 1 \\ 1 & 0 & 0 & 0 & 0 & \cdots & 0 & 1 & 1 \end{bmatrix}. \tag{A1}$$

The determinant of $A$ is given by

$$|A| = \begin{cases} 0, & K = 0 \,(\text{mod } 3), \\ -3, & K = 1 \,(\text{mod } 3), \\ 3, & K = 2 \,(\text{mod } 3). \end{cases} \tag{A2}$$

Therefore if $K \neq 0 \pmod{3}$, the inverse matrix exists. Suppose that $K \pmod{3}$ is not zero. Then the inverse mapping is expressed by $f^{-1}(w) = Bw \pmod{2}$ using a new matrix $B$. The $i \in \{0, 1, 2, \ldots, K-1\}$-th row of $B$ is defined by shifting the following bit sequence to the right-hand side $i$ times:

$$b_K = \begin{cases} (1,1,0,1,1,0,\ldots,1,1,0,\ldots), & K = 1 \,(\text{mod } 3), \\ (1,0,1,1,0,1,\ldots,1,0,1,\ldots), & K = 2 \,(\text{mod } 3). \end{cases} \tag{A3}$$

Let us consider two configurations $w = \{w_0, w_1, \ldots, w_{K-1}\}^T$ and $w' = \{w'_0, w'_1, \ldots, w'_{K-1}\}^T$. If $K = 1 \pmod{3}$ and $Bw = Bw' \pmod{2}$, then we have

$$w_0 + w_1 + w_3 + w_4 + \ldots = w'_0 + w'_1 + w'_3 + w'_4 + \ldots (\text{mod } 2), \tag{A4}$$

$$w_0 + w_1 + w_2 + w_4 + \ldots = w'_0 + w'_1 + w'_2 + w'_4 + \ldots (\text{mod } 2), \tag{A5}$$

$$w_1 + w_2 + w_3 + w_5 + \ldots = w'_1 + w'_2 + w'_3 + w'_5 + \ldots (\text{mod } 2). \tag{A6}$$

The left side of (A4)+(A5)+(A6) is $w_1$, and the right side of (A4)+(A5)+(A6) is $w_1'$. Thus we find $w_1 = w_1'$. Similarly we can show that $w_i = w_i'$ for any $i \in \{0, 1, 2, \ldots, K-1\}$. Consequently, if $f^{-1}(w) = f^{-1}(w')$ then $w = w'$. This means that the mapping $f(w)$ is an injectition. Since the inverse matrix exists in the case of $K$ (mod 3)=1 the mapping $f(w)$ is a bijection. In the same way we can show that the $f(w)$ is a bijection mapping if $K=2$ (mod 3).

If $K=0$ (mod 3) the mapping $f(v)$ is not an injection. For example, both of $\{0, 0, 0, \ldots, 0, 0, 0\}$ and $\{0, 1, 1, 0, 1, 1, \ldots, 0, 1, 1\}$ are mapped to $\{0, 0, 0, \ldots, 0\}$.

**APPENDIX B**

The aim of this appendix to show that there are eigenvectors of $M$ satisfying (22), however, we prove it in more general form.

Let $M$ be an orthogonal matrix with size $N = 2^K$, and $\hat{i} = (N-1)-i$. For $0 \le i, j \le N-1$, $\omega_{ij}$ is defined by $(-1)^{\sharp_i + \sharp_j}$, where $\sharp_i$ denotes the number of 1 of a decimal number $i$ in binary form. If there are the following relations:

$$M_{\hat{i}\hat{j}} = \omega_{ij} M_{ij}, \quad i, j = 0, 1, \ldots N-1, \tag{B1}$$

then there is an $N$-dimensional orthnormal basis satisfying condition: (1) each basis is an eigenvector of $M$,

and (2) the absolute value of the $i$-th element is the same with the absolute value of the $\hat{i}$-th element for $i = 0, 1, \ldots, N/2 - 1$.

Let $c_i$ be a complex number whose absolute value is 1 and $n = N/2$. Then from the condition (2), the $i$-th element of basis $v_j (j = 0, 1, \ldots, n-1)$ is expressed by

$$v_{ij} = \begin{cases} b_{ij}, & i = 0, 1, \ldots, n-1, \\ c_{\hat{i}} b_{\hat{i}j}, & i = n, n+1, \ldots, N-1. \end{cases} \tag{B2}$$

From the condition (1), the vector $v_j$ must be the eigenvector of $M$. If the vector $v_j$ is an eigenvector corresponding to an eigenvalue $\lambda$, then the equation $M v_j = \lambda v_j$ is satisfied.

Submitting $v_j$ in (B2) into this equation, we have

$$\sum_{k=0}^{n-1}(M_{ik} + c_k M_{i\hat{k}})b_{kj} = \lambda b_{ij}, \quad i = 0, 1, \ldots, n-1, \tag{B3}$$

$$\sum_{k=0}^{n-1}(c_k \omega_{ik} M_{ik} + \omega_{i\hat{k}} M_{i\hat{k}})b_{kj} = c_i \lambda b_{ij}, \quad i = 0, 1, \ldots, n-1. \tag{B4}$$

If (B3) and (B4) hold, the left-hand side of (B3) multiplied by $c_i$ must be the right-hand side of (B4). Suppose (B3) is satisfied and $c_i$ is the solution of the following simultaneous equations:

$$c_i - c_k \omega_{ik} = 0, \tag{B5}$$

$$c_i c_k - \omega_{i\hat{k}} = 0. \tag{B6}$$

Then (B4) is also satisfied. A possible solution of the simultaneous equation is given by

$$c_i = \begin{cases} (-1)^{\sharp i}, & K = \text{even}, \\ (-1)^{\sharp i}\sqrt{-1}, & K = \text{odd}. \end{cases} \tag{B7}$$

After submitting (B7) into (B5) and (B6), we confirm easily that $c_i$ is a solution by using the following transformation

$$(-1)^{\sharp \hat{j}} = \begin{cases} (-1)^{\sharp j}, & K = \text{even}, \\ -(-1)^{\sharp j}, & K = \text{odd}. \end{cases} \tag{B8}$$

Let us define a new matrix $H$ whose element of the $i$-th row and the $j$-th is given by $(M_{ij} + c_j M_{i\hat{j}})$ for $0 \leq i, j \leq n-1$. From (B3) we find $\{b_0, b_1, \ldots, b_{n-1}\}$ is an eigenvector of $H$. We here show that the matrix $H$ is an orthogonal matrix. Let $h_j$ and $m_j$ be the $j$-th column vector in $H$ and $M$. Furthermore let $\bar{m}_j$ be a vector $\{M_{0j}, M_{1j}, \ldots, M_{(n-1)j}\}^T$. An inner product $h_i \cdot h_j$ is given by

$$h_i \cdot h_j = (\bar{m}_i + c_i \bar{m}_{\hat{i}}) \cdot (\bar{m}_j + c_j^* \bar{m}_{\hat{j}})$$
$$= \bar{m}_i \cdot \bar{m}_j + c_i c_j^* \bar{m}_{\hat{i}} \cdot \bar{m}_{\hat{j}} + c_i \bar{m}_{\hat{i}} \cdot \bar{m}_j + c_j^* \bar{m}_i \cdot \bar{m}_{\hat{j}}. \tag{B9}$$

To express the left-hand side of (B9) using $m_i$, $m_j$, $m_{\hat{i}}$ and $m_{\hat{j}}$, we express $m_i \cdot m_j$ using $\bar{m}_i$, $\bar{m}_j$, $\bar{m}_{\hat{i}}$ and $\bar{m}_{\hat{j}}$ as

$$m_i \cdot m_j = \sum_{k=0}^{n-1} M_{ki} M_{kj} + \sum_{k=0}^{n-1} M_{\hat{k}i} M_{\hat{k}j}$$
$$= \bar{m}_i \cdot \bar{m}_j + c_i c_j^* \bar{m}_{\hat{i}} \cdot \bar{m}_{\hat{j}}. \tag{B10}$$

From (B10) we have

$$c_i m_{\hat{i}} \cdot m_j = c_i \bar{m}_{\hat{i}} \cdot \bar{m}_j + c_j^* \bar{m}_i \cdot \bar{m}_{\hat{j}}, \tag{B11}$$

where $c_i c_{\hat{i}} = 1$ is used. Comparing the right-hand of (B9) with the right-hand sides of (B10) and (B11), the inner product $h_i \cdot h_j$ is given by

$$h_i \cdot h_j = m_i \cdot m_j + c_i m_{\hat{i}} \cdot m_j, \tag{B12}$$

for $0 \leq i, j \leq N - 1$.

Since the matrix $M$ is the orthogonal matrix, $\{m_0, \ldots, m_{N-1}\}$ is an orthonormal basis. Thus the inner product is

$$h_i \cdot h_j = \begin{cases} 0, & i \neq j, \\ 1, & i = j. \end{cases} \tag{B13}$$

This means that the matrix $H$ is an orthogonal matrix. As a result, we can find an $n$-dimensional orthonormal basis $\{s_0, \ldots, s_{n-1}\}$ whose element is an eigenvectors of $H$. The inner product $v_i \cdot v_j$ is given by

$$\begin{aligned} v_i \cdot v_j &= \sum_{k=0}^{n-1} b_{ki}(b_{kj})^* + \sum_{k=0}^{n-1} c_{\hat{k}} c_{\hat{k}}^* b_{\hat{k}i}(b_{\hat{k}j})^* \\ &= 2 b_i \cdot b_j. \end{aligned} \tag{B14}$$

Accordingly it follows that an $n$-dimensional ortonormal basis whose element is an eigenvector of $M$ is obtained from (B2) by setting $b_{ij} = s_{ij}/\sqrt{2}$.

Using the same procedure we can find another basis in the form

$$\tilde{v}_{ij} = \begin{cases} b_{ij}, & i = 0, 1, \ldots, n-1, \\ -c_i b_{ij}, & i = n, n+1, \ldots, N-1. \end{cases} \tag{B15}$$

For any $i$ and $j$, an inner product $v_i \cdot \tilde{v}_j$ is always zero. Thus $\{v_0, v_1, \ldots, v_{N-1}, \tilde{v}_0, \tilde{v}_1, \ldots, \tilde{v}_{N-1}\}$ is an $N$-dimensional ortonormal basis satisfying conditions (1) and (2).

---

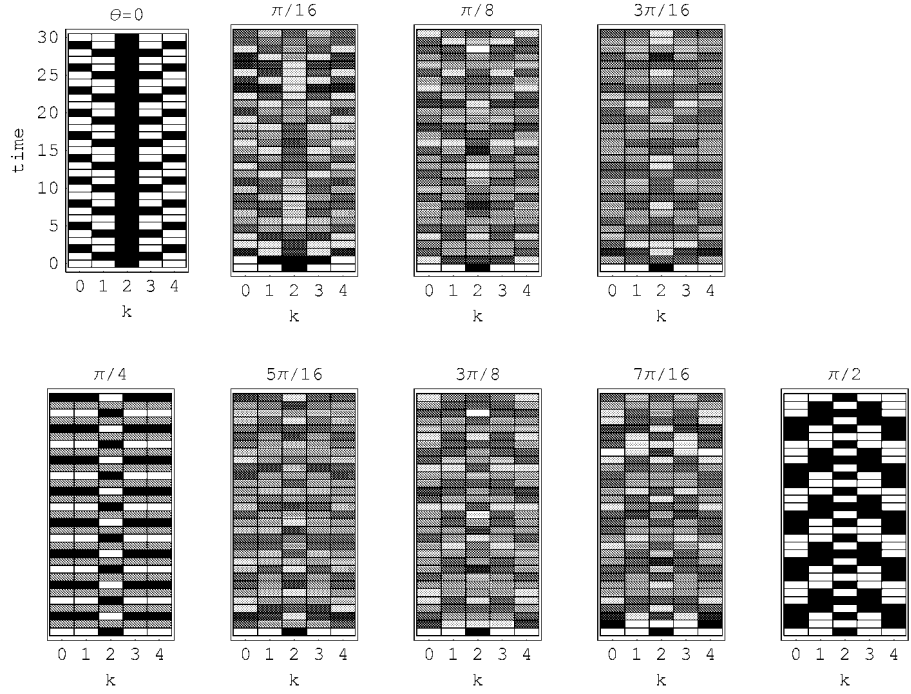

FIG. 1: Time-evolution of QCA based on CA with Wolfram's rule 150. The darkness is proportion to the probability of finding an active cell. The number of cells is five and the initial state is $\{0,0,1,0,0\}$ in common.

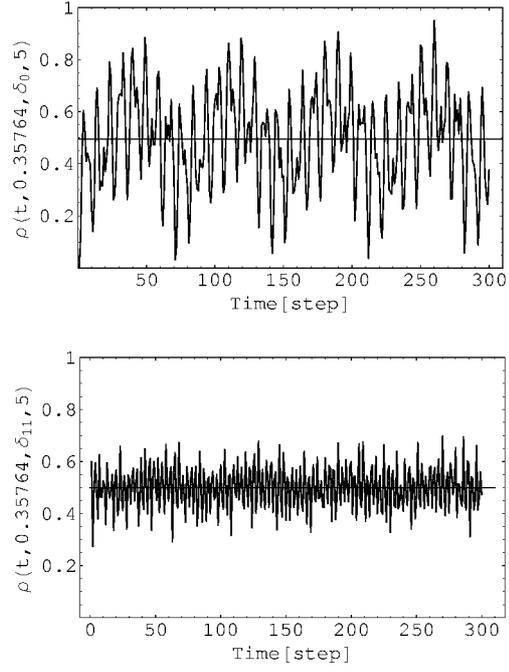

FIG. 2: The change of the mean density of active cells in QCA with $K = 5$ and $\theta = 0.35764$. The initial states are (a) $\xi_0 = \{0,0,0,0,0\}$ and (b) $\xi_{11} = \{0,1,0,1,1\}$.

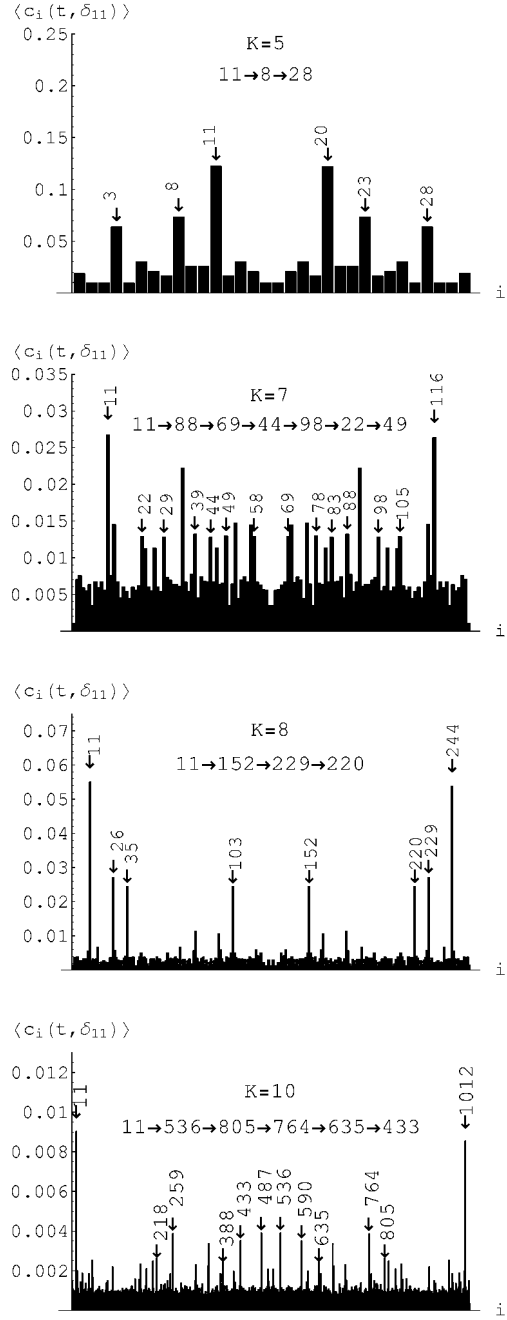

FIG. 3: The distribution of time-average probability of finding the configuration $\xi_i$ in QCA with size $K = 5, 7, 8, 10$. The parameter $\theta$ and the initial state are 0.7 and $\delta_{11}$ in common. The sequences of numbers connected with arrows indicate changes of configurations in CA.

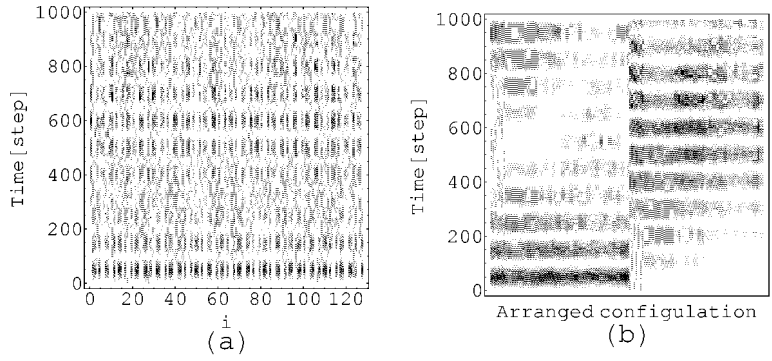

FIG. 4: (a) The time-evolution of probability of finding a configuration $\xi_i$. The darkness is proportion to the value of the probability. (b) The same simulation results with (a), but the order of configuration is arranged. The relation between the $i$-th configuration and the $(2^{K-1} + i)$-th configuration is bit-reversal.